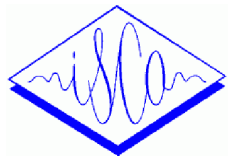

# Speaker Recognition for Children's Speech

*Saeid Safavi, Maryam Najafian, Abualsoud Hanani, Martin Russell, Peter Jančovič, Michael Carey*

School of Electronic, Electrical and Computer Engineering, University of Birmingham,
Birmingham, B15 2TT, England

`{sxs796, mxn978, m.j.russell, p.jancovic, m.carey}@bham.ac.uk, ahanani@birzeit.edu`

## Abstract

This paper presents results on Speaker Recognition (SR) for children's speech, using the OGI Kids corpus and GMM-UBM and GMM-SVM SR systems. Regions of the spectrum containing important speaker information for children are identified by conducting SR experiments over 21 frequency bands. As for adults, the spectrum can be split into four regions, with the first (containing primary vocal tract resonance information) and third (corresponding to high-frequency speech sounds) being most useful for SR. However, the frequencies at which these regions occur are from 11% to 38% higher for children. It is also noted that sub-band SR rates are lower for younger children. Finally results are presented of SR experiments to identify a child in a class (30 children, similar age) and school (288 children, varying ages). Class performance depends on age, with accuracy varying from 90% for young children to 99% for older children. The identification rate achieved for a child in a school is 81%.

## 1. Introduction

As human interaction with computers becomes more pervasive, and its applications become more private and sensitive, the value of automatic Speaker Recognition (SR) based on vocal characteristics increases.

The employment of SR technology for children could be beneficial in several application areas, including, child security and protection, and education. For instance, social networking sites are most popular with teenagers and young adults, with almost half of children aged from 8 to 17 who use the internet having set up their own profile on a social networking site [1]. An SR system that identifies a child based on his or her voice, and confirms the identity of the individual with whom the child is communicating, could be a valuable safeguard for a child engaged in social networking. Other possible applications are in education. For example, an interactive educational tutor that could identify each child in a class could automatically continue a previous lesson, adapt its content to suit the child, and log the child's responses appropriately without the child needing to go through a formal login process.

Although automatic recognition of children's speech has been the subject of considerable research effort, there is little published work on issues and algorithms related to automatic verification of a child's identity from his or her speech. For example, we do not know how increases in inter- and intra-speaker variability for children's speech [4] will affect SR performance. Variability is highest for young children, converging to adult values when children reach the age of 13. Even for young children there is some evidence that the degree of variability varies significantly between individuals [6].

It has been shown that acoustic and linguistic characteristics of children's speech are different from those of adult's [3-5]. For example, children's speech is characterized by higher pitch, and perceptually important features such as formants occur at higher frequencies [4]. Consequently, the impact of bandwidth reduction on speech recognition accuracy is greater for children's speech than for adults [6, 7]. However, we do not know the significance of different frequency bands for SR for children, although the relevant studies for adult SR have been reported [2].

The success of Gaussian Mixture Model - Universal Background Model (GMM-UBM) and GMM-Support Vector Machine (GMM-SVM) approaches to adult SR motivated us to apply these techniques to our child SR task. The distribution of acoustic feature vectors for a population of speakers, is typically captured using a UBM (a speaker-independent GMM constructed using data from a variety of speakers and background conditions) [8, 9]. Speaker dependent GMMs are then built by MAP adaptation of the UBM [10]. Alternatively, discriminative approaches such as SVMs can be used, which have been shown to obtain comparable, and in some cases better, performance than GMM based systems. The combination of GMM supervectors, comprising the stacked parameters of the GMM components, with SVMs has also been successful [11]. SR systems usually employ score normalization to cope with score variability and to simplify decision threshold tuning.

This paper presents the results of experiments in SR for children's speech and is organized as follows. Section 2 describes the OGI 'Kid's' corpus of children's speech, which is used in all experiments. Our SR systems are described in section 3, and our experiments and results are presented in section 4. Section 4.1 describes a study of the utility of the information in different frequency bands for children's SR. Results of SR experiments for narrow band limited speech show that, as in the case of adults [2], the spectrum can be usefully partitioned into 4 regions, B1 to B4, with B1, B2, B3 and B4 corresponding to frequencies below 1.13kH, 0.63kHz to 3.8kHz, 2.1kHz to 5.53kHz and 3.4kHz to 8kHz, respectively. These frequencies are between 11% and 38% higher than those for adults [2]. Speaker information is concentrated in B1, which contains the primary vocal tract resonances, and B3, which contains high-frequency speech sounds such as fricatives. The speaker information in region B2 is masked by linguistic variation. It is also noted that narrow-band SR performance is consistently poorer for young children than for older children. Section 4.2 presents the results of verification and identification experiments for different age groups of children using full bandwidth speech. The best performance is obtained using a 64 component GMM-SVM system. Finally, with educational applications in mind, we simulate the problem of recognizing a single child in a class (30 children of a similar age) or a school (288 children varying in age from 5 to 13 years). Identification accuracy for a child in a class varies from 90% for the youngest children (5-8years) to 99% for the oldest children (12 years old and above). The identification rate achieved for a child in a school is 81%.

## 2. The OGI kids' speech corpus and data description

The OGI Kids' Speech corpus [13] is a collection of spontaneous and read speech recorded at the Northwest Regional School District near Portland, Oregon. The CSLU Toolkit is used for data collection. It comprises recordings of words and sentences from approximately 1100 children. A gender-balanced group of approximately 100 children per grade from Kindergarten (5-6 year olds) through to grade 10 (15–16 year olds) participated in the collection. For each utterance, the text of the prompt was displayed on a screen, and a human recording of the prompt was played, in synchrony with facial animation using the animated 3D character "Baldi". The subject then repeated the prompt, which was recorded via a head-mounted microphone and digitized at 16 bits and 16 kHz.

Four different test sets (10 seconds per utterance) from the OGI data are used in the experiments presented in this paper.

TS1: To investigate the effect of different frequency bands on SR performance for general children's speech, 359 speakers were chosen randomly (kindergarten to $10^{th}$ grade).

TS2: To investigate the effect of different frequency bands on SR performance for speech from children of different ages, 3 different age groups were selected, each containing 288 speakers. These are AG1: kindergarten to $2^{nd}$ grade (5-8 year olds), AG2: $3^{rd}$ to $6^{th}$ grade (8-12 year olds), and AG3: $7^{th}$ to $10^{th}$ grade (12-16 year olds).

TS3: To investigate the problem of identifying a single child in a school, two 'schools' of 288 randomly chosen speakers from kindergarten to 10th grade were chosen.

TS4: To investigate the problem of identifying a single child in a class, 12 'classes' of children from 3 grade groups were chosen, each containing 30 children.

## 3. Speaker recognition systems

### 3.1. Signal Analysis

Feature extraction was performed as follows. Periods of silence were discarded using an energy-based Speech Activity Detector (SAD). The speech was then segmented into 20-ms frames (10-ms overlap) and a Hamming window was applied. The short-time magnitude spectrum, obtained by applying an FFT, is passed to a bank of 24 Mel-spaced triangular band-pass filters, spanning the frequency region from 0Hz to 8000Hz. Table 1 shows the center frequency of each filter (the cut-off frequencies of a filter are the centre frequencies of the adjacent filters).

To investigate the effect of different frequency regions on SR performance, experiments were conducted using frequency band limited speech data comprising the outputs of groups of 4 adjacent filters. We considered 21 overlapping sub-bands, where the $N^{th}$ sub-band comprises the outputs of filters $N$ to $N+3$ ($N$=1 to 21). Each set of 4 filter outputs was transformed to 4 Mel Frequency cepstral coefficients (MFCCs) and mean and variance normalization [14] was applied.

For the full bandwidth experiments the outputs of all 21 filters were transformed into 19 MFCCs.

### 3.2. Modelling

Our SR systems are based on the GMM-UBM [9, 11] and GMM-SVM [11] methods.

In the GMM-UBM approach, a UBM is built using utterances from all data in the training sets of all speakers. Speaker-dependent models are obtained by MAP adaptation (adapting means only) of the UBM, using 48 second segments of speaker-specific enrollment data. The result is one UBM and 1083 speaker-dependents GMMs (a small number of speakers for whom there was very little data were not used).

In our GMM-SVM system, the training data from each individual speaker was divided into three segments and each was used to estimate the parameters of a GMM by MAP adaptation of the UBM. The adapted GMM mean vectors are then concatenated into a supervector [11], and the speaker classes are assumed to be linearly separable in the supervector space. The supervectors are used to build one SVM for each speaker, by treating that speaker as the 'target' class and the others as the 'background' class.

In our recognition systems, the score for each speaker model is normalized using the highest score across all speakers (max-log-likelihood score normalization).

Table 1: *The Center Frequencies for 24 Mel-Spaced Band-Pass Filters.*

| FILTER NUMBER | CENTRAL FREQUENCY(HZ) | FILTER NUMBER | CENTRAL FREQUENCY(HZ) |
|---|---|---|---|
| 1 | *156* | 13 | *1843* |
| 2 | *281* | 14 | *2062* |
| 3 | *406* | 15 | *2343* |
| 4 | *500* | 16 | *2656* |
| 5 | *625* | 17 | *3000* |
| 6 | *750* | 18 | *3375* |
| 7 | *875* | 19 | *3812* |
| 8 | *1000* | 20 | *4312* |
| 9 | *1125* | 21 | *4906* |
| 10 | *1281* | 22 | *5531* |
| 11 | *1437* | 23 | *6281* |
| 12 | *1625* | 24 | *7093* |

### 3.3. Verification and Identification experiments

Verification experiments were conducted using a version of the methodology developed for the NIST speaker recognition evaluations. Each test utterance was scored against the 'true' (correct) speaker model and 10 'impostor' models. Results are presented in terms of percentage Equal Error Rate (EER), calculated using the standard NIST software. Identification experiments involved scoring each test utterance against a fixed test set of speaker models and assigning the model to the class with the highest score. Test sets TS1 to TS4 from Section 2 were used.

## 4. Experimental results and discussion

### 4.1. Experiments on isolated sub-bands

In this section, we study the effect of different sub-bands on verification and identification performance for children's speech from the OGI corpus. SR tests are conducted separately on 21 sub-bands, each consisting of four consecutive channels (Section 3.1).

Figures 1(a) and (b) show the verification and identification performances, respectively, for the 359 speaker test set (TS1) on each of the 21 sub-bands, using 64 component GMM-UBM and GMM-SVM systems (64 component GMMs were found to be adequate for these 4 dimensional sub-bands). Overall, it is clear that the GMM-SVM approach outperforms GMM-UBM.

In the case of verification, Figure 1(a) shows sub-band EERs varying between 10% and 37%. For identification (Figure 1(b)) the sub-band identification rates vary between 5% and 34%.

From Figure 1 it is evident that, as in the case of adult speech [2], it is convenient to partition the spectrum into 4 frequency regions, B1 to B4, where B1 corresponds to sub-bands 1-5 (0-1.13kH), B2 to sub-bands 6-14 (0.63kHz – 3.8kHz), B3 to sub-bands 15-18 (2.1kHz – 5.53kHz), and B4 to sub-bands 19-21 (3.4kHz to 8kHz). The most useful bands for SR are B1, which contains individual differences in the part of the spectrum due to primary vocal tract resonances and nasal speech sounds, and B3, which contains information relating to high-frequency speech sounds such as fricatives. One would expect B2 to be useful since it contains information about vocal tract resonances, however the speaker specific information in this region appears to be masked by variations due to the linguistic content of the signal. Interestingly, the GMM-SVM system is able to extract more speaker-specific information from B2 than the GMM-UBM system. The importance of fricatives (and hence region B3) for SR has been noted previously in [15]. Frequency regions similar to B1 to B4 were identified in [2] for adult SR on TIMIT. However, compared to the adult values, the frequency ranges spanned by these bands for children's speech are increased by approximately 38% (B1), 21% (B2) and 11% (B3).

Figure 2 shows sub-band speaker identification rates for three different age-groups of children, namely AG1, AG2 and AG3 (described in section 2). The figure shows that in almost all cases the best performance is obtained for the older children, and identification rate decreases for younger children. The figure shows the same fall in performance between B1 and B2, and increase between B2 and B3, for all three age groups. However, one would expect these changes to take place at higher frequencies for younger children, since in general younger children have shorter vocal tracts. Close inspection of figure 2 indicates that this is the case.

The result for the oldest children (7th to 10th grade, AG3) is consistent with published result for adult speaker identification on TIMIT [2].

### 4.2. Full-bandwidth SR for children's speech

Table 2 shows the results of SR experiments on full-bandwidth speech for the three age groups of children (AG1 to AG3, 288 children per group), using 1024 component GMM-UBM and GMM-SVM systems and a 64 component GMM-SVM system.

The choice of 1024 components for the GMM-UBM system was made empirically on a separate evaluation set. Both identification rate and EER improve as the ages of the children increase. For example the EER falls by 70% from 2.1% for the youngest to 0.64% for the oldest children. The corresponding increase in identification rate is 38%. The performance of the 1024 component GMM-SVM system was unexpectedly poor. An experiment on a separate evaluation set showed that the best number of GMM components for this system is 64, due to the short test utterances. The performance of the 64 component GMM-SVM system is shown in column 4 of Table 2. Verification performance is similar to that obtained for the 1024 component GMM-UBM system, but the identification rates are between 9% and 20% better for the 64 component GMM-SVM system.

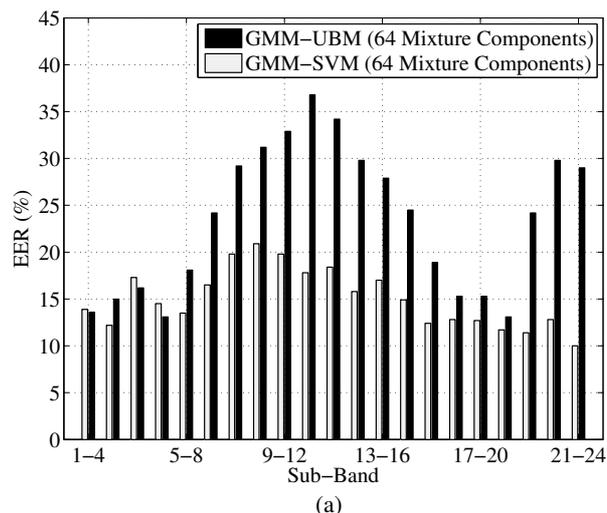

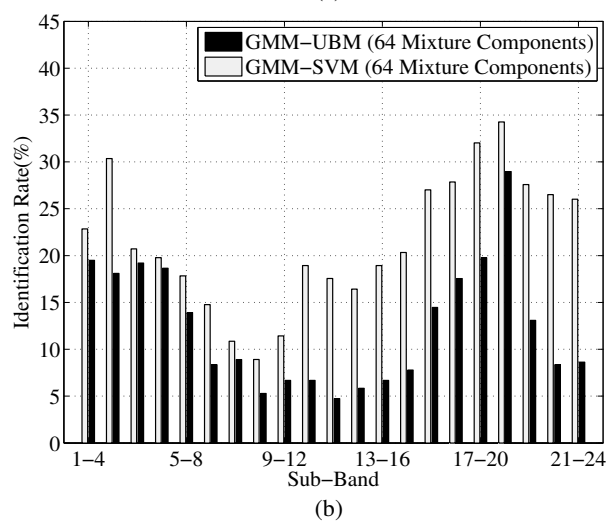

Figure 1: *Sub-band speaker verification rate (EER) (a), and speaker identification rate (b) for child speech from OGI corpus for different frequency bands.*

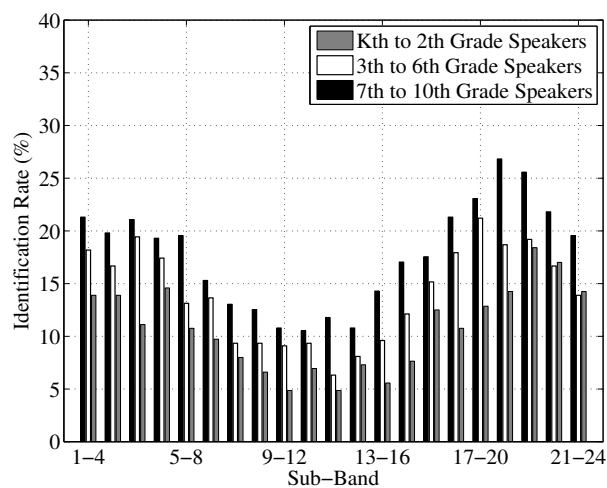

Figure 2: *Sub-band speaker identification rates for three age groups of children, namely AG1, AG2 and AG3.*

Table 2: *SR performance for three different grade groups (AG1, AG2 and AG3).*

|  | GMM-UBM(1024) | GMM-SVM(1024) | GMM-SVM(64) |
|---|---|---|---|
| **Verification** | EER (%) | EER (%) | EER (%) |
| **AG1(K-2)** | 02.10 | 06.94 | **02.00** |
| **AG2(3-6)** | 01.33 | 03.48 | **01.21** |
| **AG3(7-10)** | **00.64** | 02.83 | 00.84 |
| **Identification** | ID (%) | ID (%) | ID (%) |
| **AG1(K-2)** | 62.15 | 38.54 | **75.00** |
| **AG2(3-6)** | 80.56 | 79.17 | **88.19** |
| **AG3(7-10)** | 85.71 | 83.33 | **93.06** |

The purpose of our final experiment is to evaluate SR performance for children's speech on tasks which are representative of potential applications. Table 3 shows the results of using the 64 component GMM-SVM system to recognize an individual child in a class (30 children from the same grade group as the target child) or school (288 children uniformly distributed across grades). The 'class' experiment is conducted for simulated classes from age groups AG1, AG2 and AG3. For each age group, the experiment was repeated for 4 random simulated classes, and the average result is given in Table 2.

The results show that a child in a class is identified with accuracies of approximately 90%, 96% and 99% for classes of 30 children in age groups AG1, AG2 and AG3, respectively. As with speech recognition, speaker recognition appears to be significantly more difficult for younger children.

The identification rate for an individual child in a school of 288 children is 81%.

Table 3: *SR accuracy for identifying a child in a class of school.*

|  | GMM-SVM (64) | |
|---|---|---|
| **SR Performance** | EER (%) | ID (%) |
| **Classroom (AG1)** | 01.92 | 89.99 |
| **Classroom (AG2)** | 01.04 | 95.83 |
| **Classroom (AG3)** | 00.83 | 99.16 |
| **School (Kth-10th)** | 01.74 | 81.00 |

## 5. Conclusions

This paper presents the results of experiments in SR for children's speech. A study of the utility of different narrow frequency bands for child SR has shown that, as with adults, the spectrum can be usefully partitioned into 4 regions, referred to as B1 to B4, such that most useful speaker information is concentrated in B1, which contains the primary vocal tract resonances, and B3, which contains high-frequency speech sounds such as fricatives. However, the frequencies at which these regions occur are between 11% and 38% higher for young children than for adults. It has also been shown that sub-band SR identification rates are consistently poorer for younger children than for older children.

Experiments which simulate recognition of an individual child in a class or a school, using a 64 component GMM-SVM system, show that identification rates for a child in a class vary between 90% for the youngest to 99% for the oldest children, and that the identification rate for a child in a school is 81%.

## 6. References

[1] Anonymous "Engaging with social networking sites," vol. 2011.

[2] L. Besacier, J. Bonastre and C. Fredouille, "Localization and selection of speaker-specific information with statistical modeling," Speech Commun., vol. 31, pp. 89-106, 2000.

[3] S. Nittrouer and D. Whalen, "The perceptual effects of childN adult differences in fricative-vowel coarticulation," J. Acoust. Soc. Am., vol. 86, pp. 1266-1276, 1989.

[4] S. Lee, A. Potamianos and S. Narayanan, "Acoustics of children's speech: Developmental changes of temporal and spectral parameters," J. Acoust. Soc. Am., vol. 105, pp. 1455-1468, 1999.

[5] M. Gerosa, S. Lee, D. Giuliani and S. Narayanan, "Analyzing children's speech: An acoustic study of consonants and consonant-vowel transition," Proc. IEEE Int. Conf. Acoustics, Speech and Signal Processing, 2006. ICASSP 2006, pp. I-393-I-396, 2006.

[6] Q. Li and M. J. Russell, "An analysis of the causes of increased error rates in children's speech recognition," in Seventh International Conference on Spoken Language Processing, 2002.

[7] S. Yildirim, S. Narayanan, D. Boyd and S. Khurana, "Acoustic analysis of preschool children's speech," in Proc. 15th ICPhS, 2003, pp. 949–952, 2003.

[8] R. C. Rose and D. A. Reynolds, "Text independent speaker identification using automatic acoustic segmentation," Proc. IEEE Int. Conf. Acoustics, Speech and Signal Processing. ICASSP-90, pp. 293-296, 1990.

[9] D. A. Reynolds, T. F. Quatieri and R. B. Dunn, "Speaker verification using adapted Gaussian mixture models," Digital Signal Processing, vol. 10, pp. 19-41, 2000.

[10] D. A. Reynolds, "Speaker identification and verification using Gaussian mixture speaker models," Speech Commun., vol. 17, pp. 91-108, 1995.

[11] W. Campbell, D. Sturim, D. Reynolds and A. Solomonoff, "SVM based speaker verification using a GMM supervector kernel and NAP variability compensation," Proc. IEEE Int. Conf. Acoustics, Speech and Signal Processing. ICASSP'2006 , 2006.

[12] F. Bimbot, J. F. Bonastre, C. Fredouille, G. Gravier, I. Magrin-Chagnolleau, S. Meignier, T. Merlin, J. Ortega-García, D. Petrovska-Delacrétaz and D. A. Reynolds, "A tutorial on text-independent speaker verification," *EURASIP Journal on Applied Signal Processing,* vol. 2004, pp. 430-451, 2004.

[13] K. Shobaki, J. P. Hosom and R. A. Cole, "The OGI kids' speech corpus and recognizers," in Sixth International Conference on Spoken Language Processing, 2000 .

[14] J. Pelecanos and S. Sridharan, "Feature warping for robust speaker verification," in Proc. Speaker Odyssey, 2001.

[15] E. S. Parris and M. J. Carey, "Discriminative phonemes for speaker identification," in Third International Conference on Spoken Language Processing, 1994.